\ifx\mnmacrosloaded\undefined \input mn\fi
\newif\ifAMStwofonts
\ifCUPmtplainloaded \else
  \NewTextAlphabet{textbfit} {cmbxti10} {}
  \NewTextAlphabet{textbfss} {cmssbx10} {}
  \NewMathAlphabet{mathbfit} {cmbxti10} {} 
  \NewMathAlphabet{mathbfss} {cmssbx10} {} 
  \ifAMStwofonts
    \NewSymbolFont{upmath} {eurm10}
    \NewSymbolFont{AMSa} {msam10}
    \NewMathSymbol{\upi}     {0}{upmath}{19}
    \NewMathSymbol{\umu}     {0}{upmath}{16}
    \NewMathSymbol{\upartial}{0}{upmath}{40}
    \NewMathSymbol{\leqslant}{3}{AMSa}{36}
    \NewMathSymbol{\geqslant}{3}{AMSa}{3E}

  \else
    \def\umu{\mu}
    \def\upi{\pi}
    \def\upartial{\partial}
  \fi
\fi
\pageoffset{-2.5pc}{0pc}
\loadboldmathnames
\pagerange{1--7}    
\pubyear{1996}
\volume{226}
\begintopmatter  

  \title{An optical counterjet in 3C66B?}

   \author{
D. Fraix-Burnet
          }


   \affiliation{
Laboratoire d'Astrophysique UMR 5571,
Observatoire de Grenoble,
BP 53, F-38041 Grenoble C\'edex 9, France,
fraix@gag.observ-gr.fr
             }


\acceptedline{Accepted September 1996 . Received July 1996 ;
  in original form 1996 July 1}

\abstract {Long exposure observations at the Canada-France-Hawaii Telescope 
of 3C66B in the I filter are presented. After subtraction of the galactic 
background, optical emission on the counterjet side is detected in 10 knots
coincident with the radio counterjet. Their radio-to-optical spectral indices
(0.5--0.6) are typical of synchrotron emission in extragalactic jets, so that
these knots possibly are the optical counterparts of the radio 
counterjet. If this is confirmed, 3C66B would be the first double-sided 
extragalactic optical jet. The optical counterjet would also be brighter
than what is predicted from the relativistic beaming interpretation of 
brightness asymmetry between the two jets. This would thus prove that the 
radiation properties are intrinsically different in the two jets.
Alternatively, these knots could have nothing to do with the counterjet. But
it seems that in this case, the optical counterjet would be fainter than
expected from the relativistic beaming interpretation, favouring intrinsic
asymmetry as well. In addition, two new optical components are found in the 
jet. 
}

      \keywords{methods: observational --
                galaxies: individual: 3C66B --
                galaxies: jets
               }
   \maketitle

%

\section{Introduction}

The spectrum of extragalactic jets is a pure synchrotron continuum, which
is essentially a power-law. This characteristic does not facilitate the
determination of the material velocity within the jet. Motions detected
at the parsec-scale close to the nucleus and also now at kpc-scales in 
M87 (Biretta, Zhou \& Owen 1995) are the only means of measuring velocities. 
This gives however only the speed of radiating particles, not 
that of the bulk of the jet. 
The two have to be distinguished because they represent two different
physics (Sol, Pelletier \& Ass\'eo 1989; Pelletier
\& Roland 1986, 1988; Fraix-Burnet \& Pelletier 1991; Fraix-Burnet 1992; 
Despringre \& Fraix-Burnet 1996).

Brightness asymmetries between jets and their counterjets in radiosources are 
generally believed to be apparent and due to Doppler effects. But
obviously, there are intrinsic structural differences in a lot of 
double-sided jets, so that this relativistic interpretation can be questioned. 
From theoretical and observational evidences, 
Fraix-Burnet (1992) argued that the jet asymmetries in radiosources are mainly
due to intrinsic differences in radiation properties. To distinguish between
these two interpretations, one should compare the physics of the jet and
counterjet. This is however very difficult with a power-law continuum 
spectrum. The spectral index is Lorentz invariant, so that it should be the
same in the two jets. A check of this point would require observations at
three frequencies, including millimetre or even sub-millimetre data, 
i.e. away enough from the break frequency which is not Lorentz invariant. 

The cutoff frequency of the synchrotron spectrum is a second characteristic 
which is very sensitive to local physical conditions within the jet. 
It is unknown for nearly all extragalactic jets 
(being in the sub-mm or far-infrared domain) except for the very few cases
where the synchrotron radiation is seen in the optical: M87, 3C273, 3C66B,
PKS0521-36, 3C264 (Crane et al. 1993), and 3C78 (Sparks et al. 1995). 
The precise shape of the cutoff is known for only the three brightest: M87
(Biretta, Stern \& Harris 1991; Meisenheimer, R\"oser \& Schl\"otelburg 
1996),
3C273 (Fraix-Burnet \& Nieto 1988) and 3C66B (Fraix-Burnet, Golombek \& 
Macchetto 1991). All known optical
jets are one-sided both in the optical and in the radio, with the
notable exception of 3C66B which shows a counterjet in the radio  
(Leahy, J\"agers \& Pooley  1986; Hardcastle et al. 1996).
If the relativistic interpretation for asymmetry is correct, the counterjet
should also radiate in the optical and should have the same intrinsic
spectrum. By deriving the Doppler factor from the
jet to counterjet intensity ratio in the radio, it is then possible to 
compute the Doppler shifted spectrum of the counterjet from the observed
spectrum of the jet. 3C66B is the only source for which this is feasible.
At a distance of 86~Mpc (for $H_0=75$~km~s$^{-1}$~Mpc$^{-1}$), 1 arcsec is 
equivalent to 417~pc in this object. 

The present study aims at detecting the optical counterjet in order to 
discriminate between the two interpretations for brightness asymmetry
and to help in understanding why so few jets are seen at optical wavelengths. 
After a presentation of the observations (Sect. 2) and a description of the
results (Sect. 3), a discussion on whether the detected structures belong to 
the jet will be made in Sect. 4, together with a comparison of the present
results with the prediction of the relativistic interpretation.

   \beginfigure*{1}
\vskip 11 cm
\caption{{\bf Figure 1.}
Residues after subtraction of elliptical models of 3C66B and its companion.
The square field is 100 arcsec wide, and cuts are -0.9 (white) to 2.5 (black)
$\mu$Jy/arcsec$^2$. Overlaid are contours of the 8~GHz 1.25 arcsec 
resolution map from Hardcastle et al. (1996).
              }
\input psfig.sty
\includegraphics{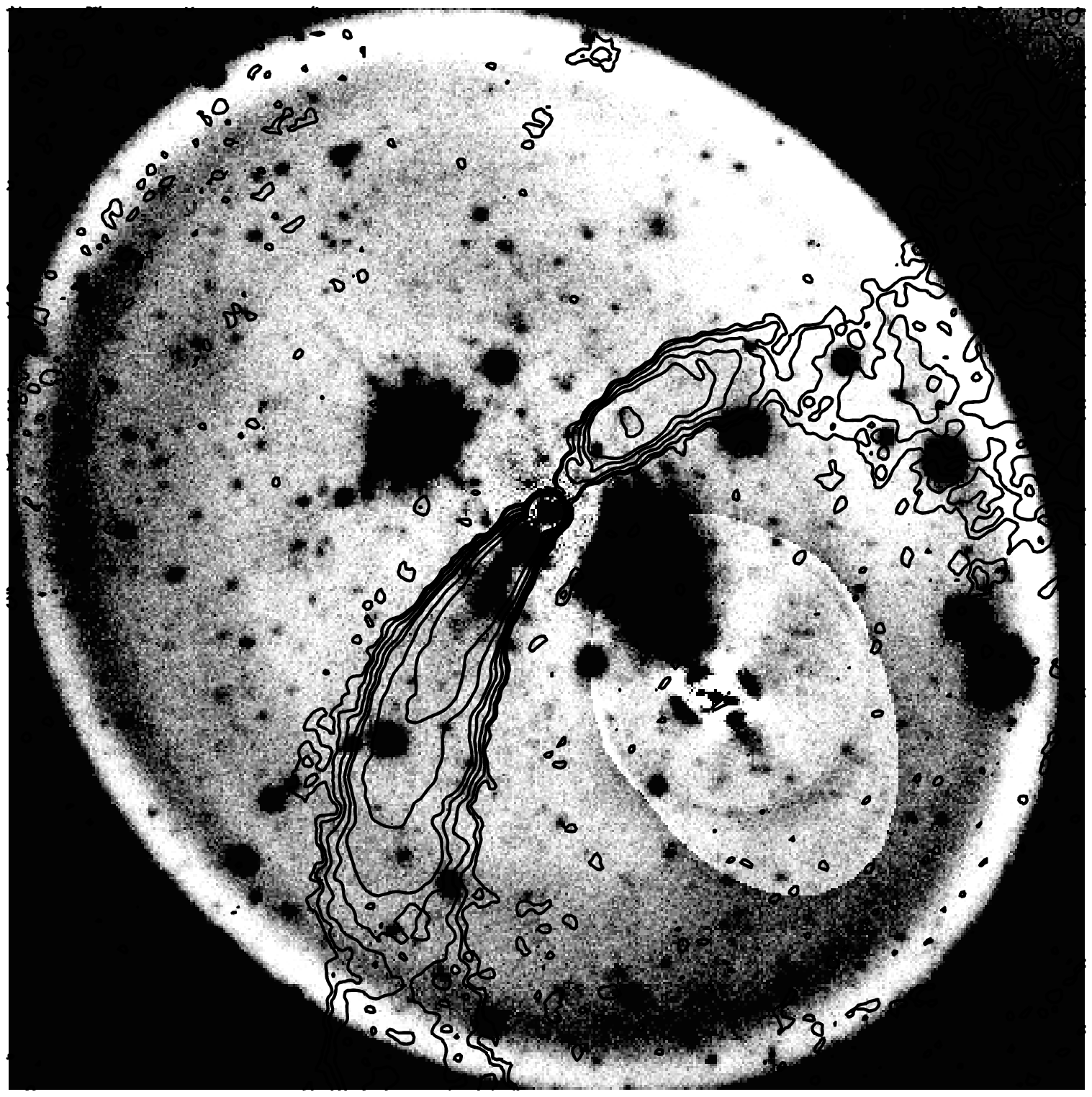}
    \endfigure

\section{Observations and data reduction}

Observations were made at the Canada-France-Hawaii Telescope (CFHT) on
September 24-27 and November 23-29, 1995. The detector was MOCAM, but
only one of the four 2048$\times$2048 CCDs (chip U2) was used. The scale
of 0.204 arcsec per pixel provides a 7 arcmin field of view. The galaxy 3C66B 
was placed near the upper (North) border of the chip so that the very
bright star 1.3 arcmin to the North was out of the CCD avoiding saturation
spikes over the galaxy. To avoid non-linearity at high exposure levels on
the center of 3C66B (for galactic background modelling and photometric
purposes), the exposure time was limited to 10 minutes, except for 3 exposures
taken through clouds where the integration time was set to 20 minutes. 
85 images through the I filter were obtained totalling nearly 15 hours of 
exposure time.  
The FWHM of the images varied from 0.75 to 1.6 arcsec with a strong 
concentration between 0.8 and 1.2 arcsec. The data were divided in two equal 
groups above and below FWHM$=1$ arcsec\ (respectively 43 and 42 images). The 
medians of the high-resolution and low-resolution groups have FWHM of 0.90 and 
1.14  arcsec respectively, while the total median of all images have 
FWHM$=1.01$ arcsec.  

Elliptical modelling of the galaxy 3C66B and its companion (at 23 arcsec 
to the SE) were performed with the ISOPHOTE package within IRAF. This task is 
complicated by a 13.7 I-magnitude star at 14 arcsec to the NNW of 3C66B.
The result of the subtraction of this
elliptical model (up to a radius of 40 arcsec) is shown in Fig.~1. 
The modelling has been
optimized so that the background around the jet and counterjet is as flat 
as possible. As can be readily seen in Fig.~1 and
Fig.~2, there are 3 
regions where the residues reveal an excess over this elliptical isophotal
shape of the galaxy (the dark patch 14 arcsec to the NNW is the 13.7 
I-magnitude star). There is a big region between the centers of 3C66B and
its companion. This might be the clear signature of the interaction 
between both galaxies. Two arc-like regions are visible on the 
jet at about 4 and 9 arcsec from the nucleus. These three regions could
be removed in the elliptical fitting process, but with the appearance of
strong negative regions. I thus consider that they are real regions of the
galaxy that depart from elliptical modelling
of 3C66B. No further attempt to remove them was done
because they do not intervene in the counterjet region, the main
goal of this work. The companion removal has not been optimized also as can be
seen on Fig.~1 and has been done essentially to ease the elliptical fitting
of 3C66B and to get a flat background around the jet and counterjet. This
background is however very slightly negative in some places.

Flux calibration of the data used the photoelectric aperture photometry by
Keel (1988): $M_I=13.18 \pm0.02$ within a 21 arcsec radius. The same
measurement was performed on the total, high- and low-resolution median images
to obtain the calibration constant. The noise level was then
estimated to be $0.02~\mu$Jy per pixel on the total median image that will
be considered in the following.

\section{Results}

   \beginfigure*{2}
\vskip 11 cm
\caption{{\bf Figure 2.}
Center 37 arcsec square field of Fig.~1
showing new features that can be related to the jet and counterjet
(white: 0 and black: 1.5~$\mu$Jy/arcsec$^2$).
              }
\includegraphics{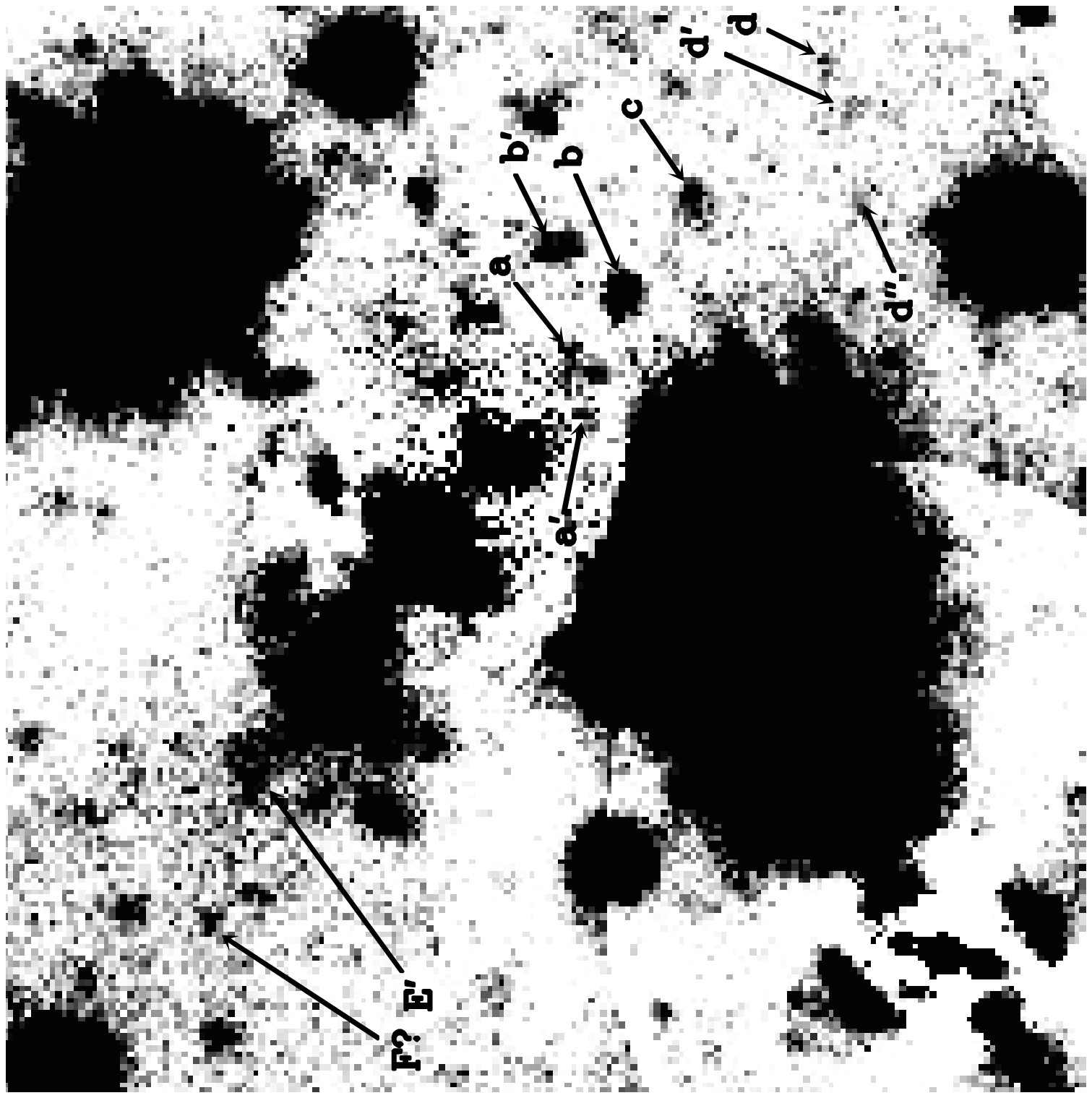}
    \endfigure
   \beginfigure*{3}
\vskip 10.5 cm
{\caption{{\bf Figure 3.}
Same as Fig.~2 (white: -0.9 and black: 1.5~$\mu$Jy/arcsec$^2$) 
with radio contours added (levels: 0.05, 0.1, 0.2,...,, 1, 2,..., 5~mJy/beam).
              }}%
\includegraphics{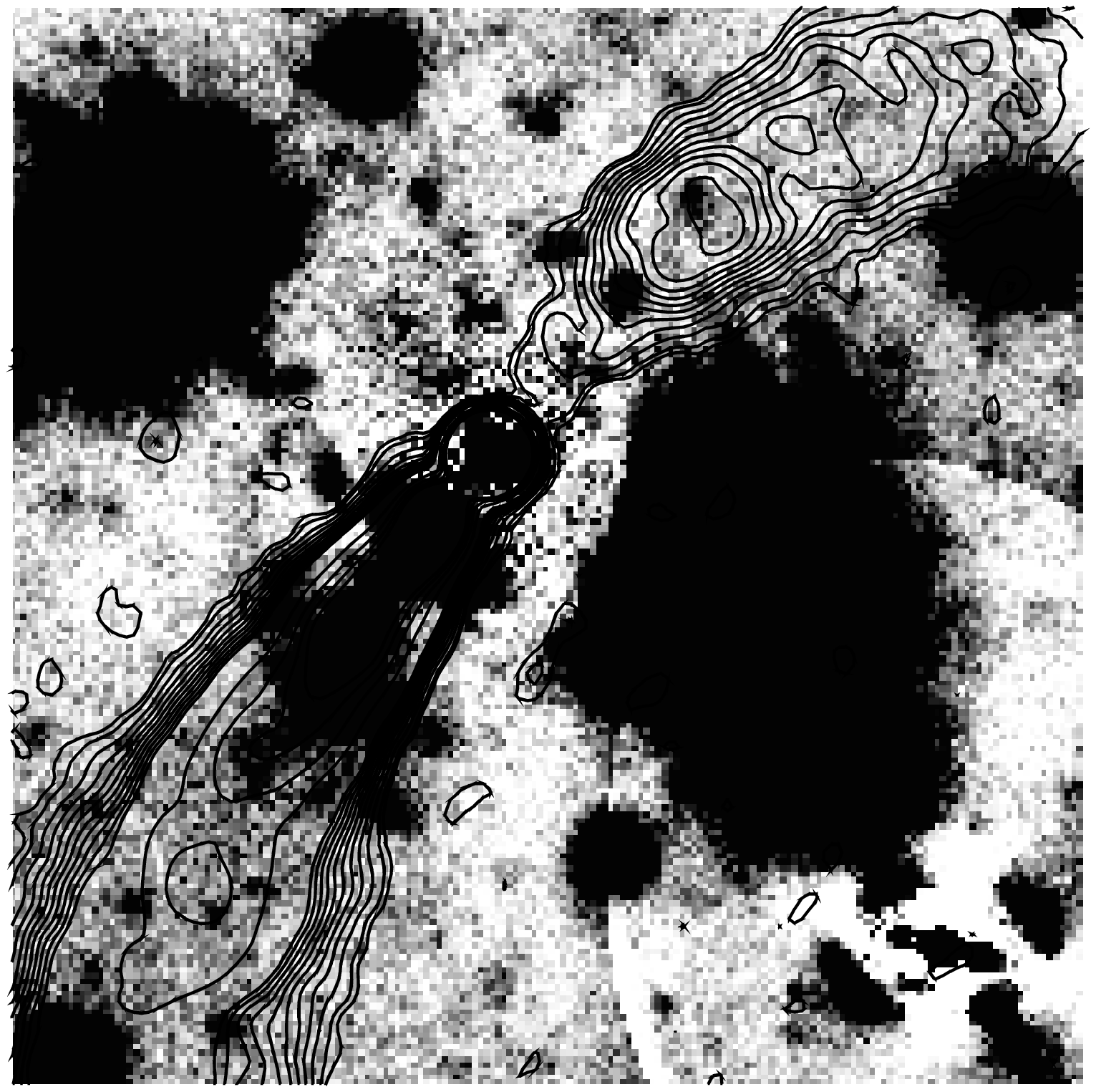}
    \endfigure

   \begintable{1}
      \caption{{\bf Table 1}
Knots characteristics: $d$ is the distance from the center of
3C66B in arcsec, PA is the position angle in degrees, $I$ is the total 
intensity ($\mu$Jy) integrated over a circular area of radius $r$ (arcsec), 
and the radio intensity ($\mu$Jy) is
measured from the 1.25 arcsec resolution 8~GHz map by Hardcastle et al. (1996)
within the same area. The spectral index is computed from the radio and $I$
intensities.}
\halign{#&\hfil#\quad&#&\hfill#\quad&\hfill#\hfill\quad&\hfil#&\hfil\quad#\hfil\cr            
            \noalign{\smallskip}
Knots & $d$\hfil &\hfil PA\hfil & $I$\hfill&$r$ & Radio\hfil &Spectral\cr
      &          &         &              &         &intensity\hfil& index \cr
            \noalign{\smallskip}            
\multispan7{\it Counterjet:}\cr
            \noalign{\smallskip}
  a    & 4.5  & 47 + 180  & 1.4 & 0.8 & 244 & 0.48 \cr
  a'   & 3.6  & 23 + 180  & 0.3 & 0.4 &  -- &  \hfil --  \cr
  b    & 7.3  & 52 + 180  & 3.6 & 0.8 & 659 & 0.48 \cr
  b'   & 7.1  & 73 + 180  & 2.3 & 0.8 &  95 & 0.35 \cr
  c    & 11.3 & 53 + 180  & 1.2 & 0.6 & 642 & 0.58 \cr
  d    & 17.7 & 50 + 180  & 0.3 & 0.4 & 150 & 0.59 \cr
  d'   & 17.1 & 44 + 180  & 0.2 & 0.4 & 167 & 0.61 \cr
  d''  & 15.3 & 34 + 180  & 0.3 & 0.4 & 151 & 0.59 \cr
            \noalign{\smallskip}
\multispan7{\it Jet:}\cr
            \noalign{\smallskip}
  A    & 1.3  & 51        & 1.6 & 0.4 & 4768 & 0.75 \cr
  B    & 2.9  & 51        &18.8 & 0.6 &11461 & 0.60 \cr
  C    & 4.1  & 53        &16.2 & 0.6 & 8892 & 0.59 \cr
  D    & 5.8  & 51        & 6.5 & 0.6 & 5518 & 0.63 \cr
  E    & 7.4  & 53        & 7.2 & 0.6 & 5409 & 0.62 \cr
  E'   & 13.2 & 52        & 1.5 & 0.6 & 2464 & 0.69 \cr
  F    & 18.3 & 59        & 0.5 & 0.4 &  916 & 0.71 \cr
}
   \endtable

The median of the 85 images is shown on Fig.~1.
The outermost isophotes are uncertain because of the
proximity of the edge of the CCD to the North. The jet is clearly visible and
overlaid contours of the radio galaxy (Hardcastle et al. 1996)
illustrate the relative sizes of the
optical jet and the two radio jets. A lot of small patches are seen which  
could be foreground stars, but also globular clusters or background galaxies
because of their somewhat extended shape.

\subsection{On the counterjet side}

A closeup view of the optical jet and counterjet regions is presented in
Fig.~2 and 3. Labelling of new potentially interesting components is
given in Fig.~2. Their interest is largely justified in Fig.~3 where
the radio contours are overlaid. Knots a, b, c, d, d' and d'' are spatially 
coincident with radio features, while knots b' could 
well be related to the jet. Knot a' will not be considered further on in this
paper since it has no radio counterpart even though it is extremely close to 
the radio jet. Of course, numerous foreground stars are present
throughout the field of the CCD and a few `knots' can be seen just North
of the counterjet. It is difficult to distinguish the labelled knots
with the others, and except maybe for knot c, they do not seem to be stellar.
There is no clear excess of such `knots' in the region of the counterjet
as compared with the 7 arcmin field of the entire CCD, as can also be seen
from Fig. 1. Thus, one cannot rule out chance coincidence of totally
unrelated objects.
Positions and intensities of these knots are given in Table~1. The spectral
indices are similar for all these knots (except b') and are slightly lower 
than those found in the jet. Note that the same spectral indices (within 0.02)
are found with the 0.75 arcsec resolution map by Hardcastle et al. (1996).

\subsection{The jet}

   \beginfigure*{4}
\vskip 8.5 cm
{\caption{{\bf Figure 4.}
{\it Left:} Image of the jet overlaid by radio contours of the 0.75 arcsec
resolution map by Hardcastle et al. (1996) (levels: 0.05, 0.1, 0.2,...,
1.0, 1.25, 1.5, 2, 3, 4, 5, 7, 10~mJy/beam). The square field is 
20 arcsec wide. Knots A, B, C, D, E have been labelled by Fraix-Burnet et al. 
(1989) and F is a radio knot defined by Hardcastle et al. (1996). 
Knot E' is a new optical knot
found in this paper that corresponds to a very distinct radio feature.
{\it Right:} Optical contours (levels: 
0.03, 0.06, 0.09, 0.13, 0.19, 0.25, 0.32, 0.38, 0.51, 0.63, 0.76
$\mu$Jy/arcsec$^2$) with the same labelling as previously.
Note the similarity between the optical contours and the radio ones
on the figure to the left.
               }}%
\includegraphics{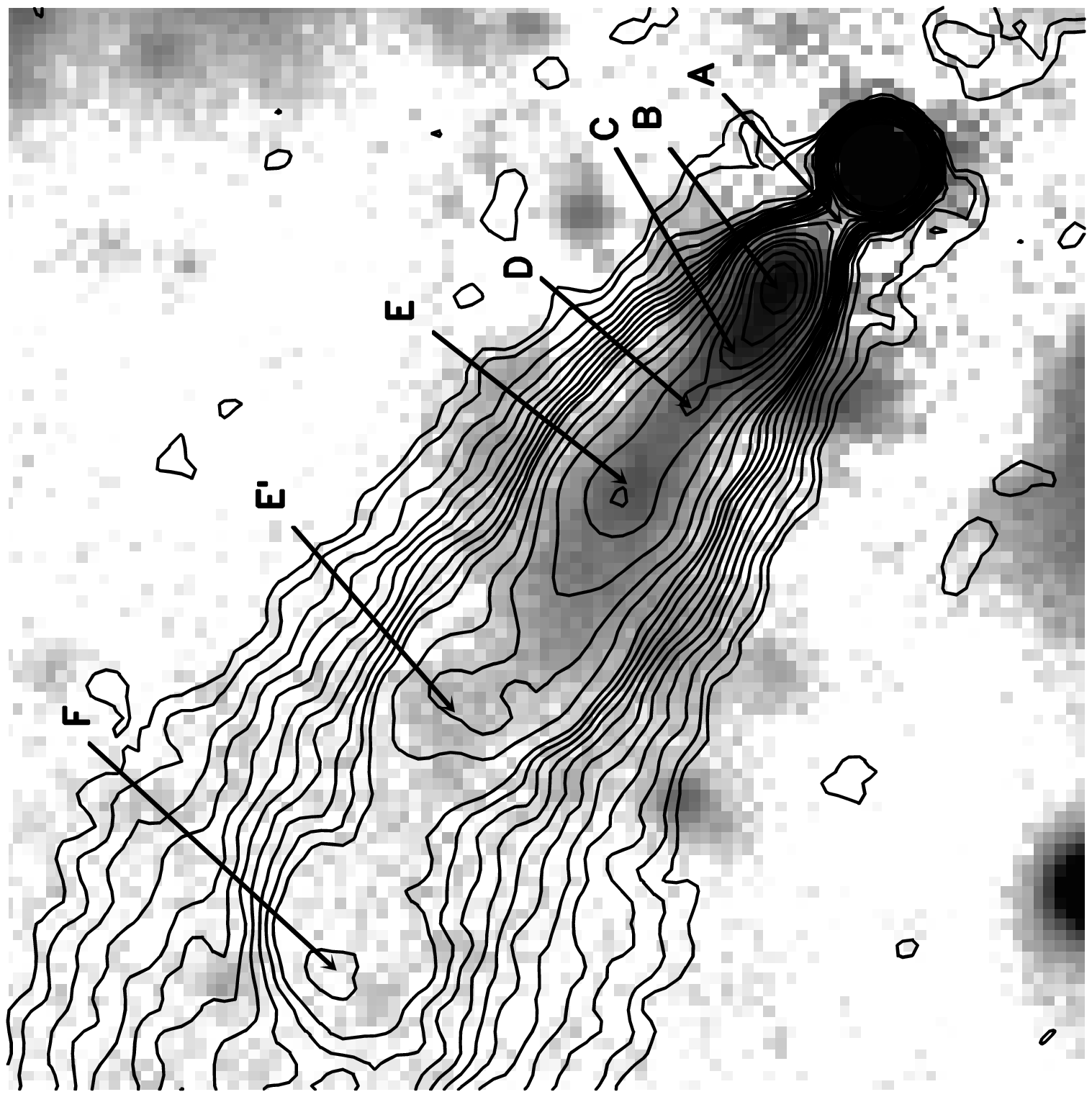}
\includegraphics{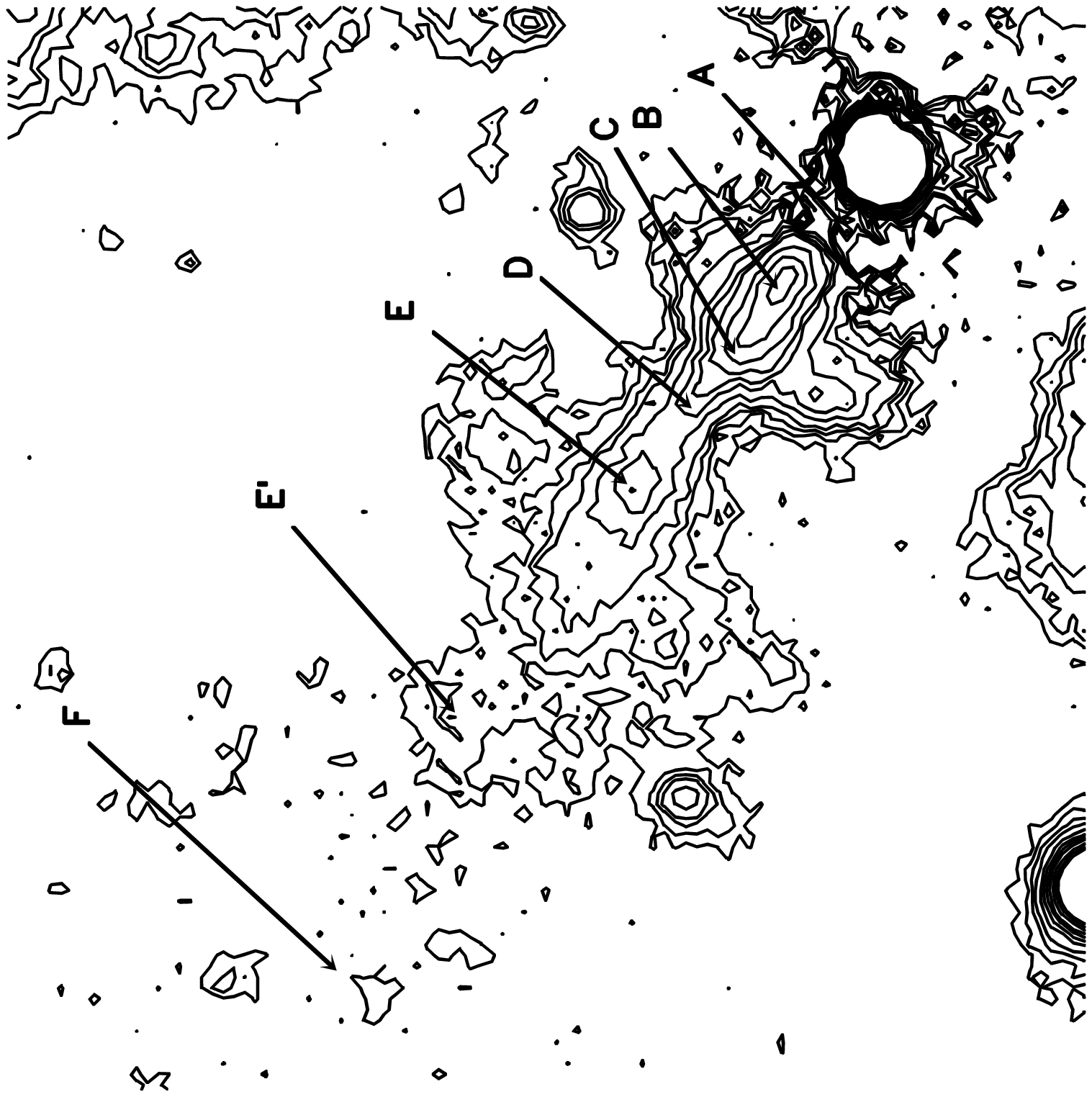}
    \endfigure

Two features newly seen in the optical are labelled on Fig.~2 
and described in Table~1. Knot E' is a new optical component of the jet, as 
can be clearly seen on Fig.~4.
Knot F has been identified from the radio by Hardcastle et al. (1996)
and there is an optical component which is here tentatively identified with it.
However, as in the case of the counterjet, it is impossible from the
present study to assert that this component is really the optical 
counterpart of radio knot F.

The similarity of contours in the radio and optical is obvious on
Fig.~4. This is confirmed by the spectral index (Table~1) being
essentially constant (excepting knot A) up to about 8 arcsec from the nucleus 
(Jackson et al. 1993). Knots E' and F have a slightly higher spectral index.

\section{Discussion}


   \beginfigure{5}
\vskip 7 cm
{\caption{{\bf Figure 5.}
The effect on the synchrotron spectrum (of spectral index $\alpha$) 
of a Doppler shift between the jet and the counterjet. The dotted line
gives the observed radio-to-optical spectral index ($\alpha'>\alpha$) of the 
jet, and the expected one ($\alpha''>\alpha'$) for the counterjet 
following the relativistic 
interpretation. Symbols `I' and `U' stand for the corresponding filters.
$\delta=3+\alpha$ or $2+\alpha$ depending on the jet model. D is the relative
Doppler factor.              }}%
\includegraphics{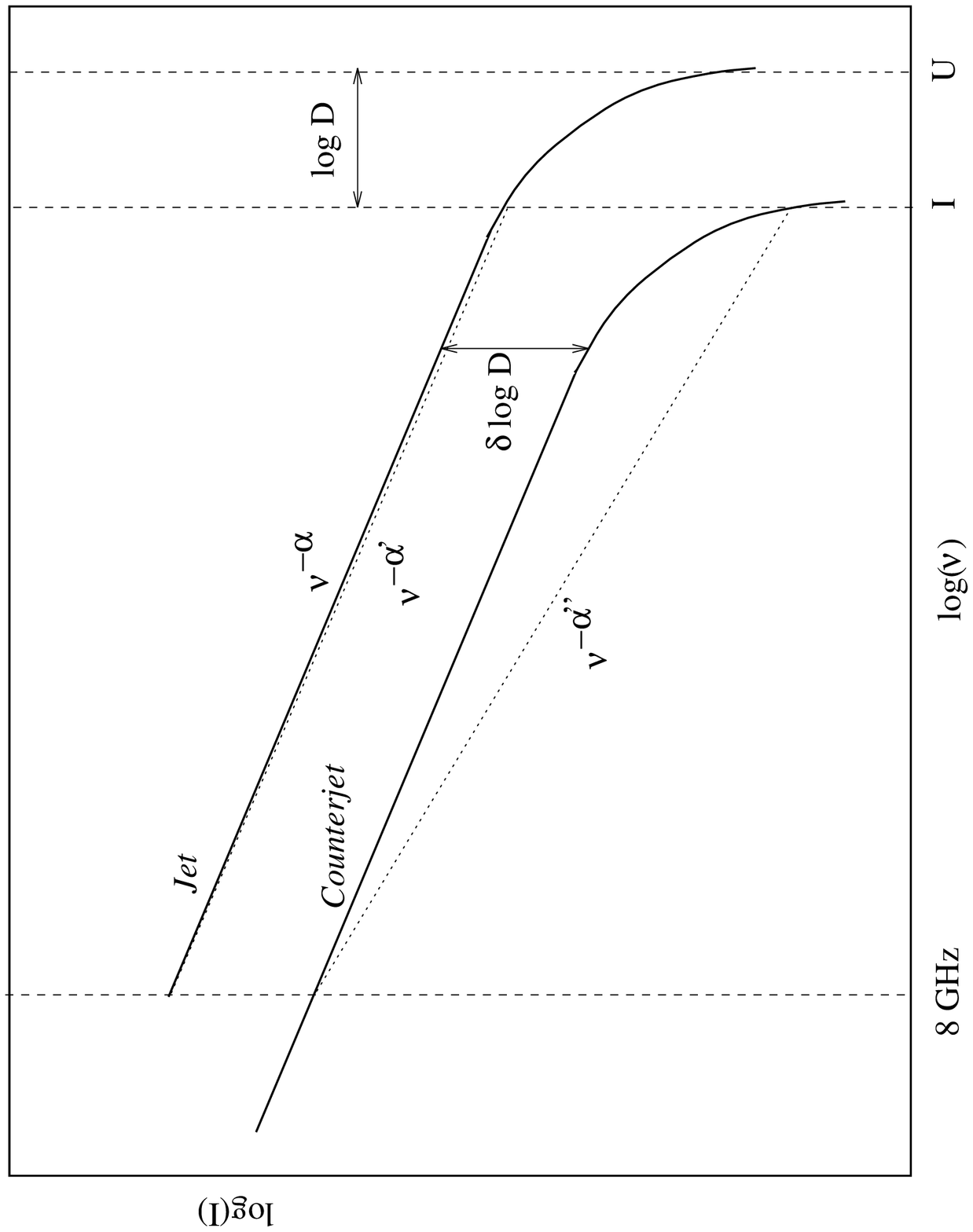}
    \endfigure

With observations in only one filter, it is difficult to determine whether the
identified knots belong to the counterjet. They do not appear stellar, 
except may be for knot c. Comparison between elliptical gaussian fits 
of these knots made on the
high-resolution and the low-resolution median images do not point to a stellar
nature. The spectral indices (between 0.5 and 0.6) derived by assuming they 
belong
to the counterjet (Table~1) are consistent with synchrotron radiation 
generally found in jets. Only knot b' departs from this
figure. It is thus entirely plausible that at least a few of these knots
are the optical counterparts of the radio jet.

Let us compare our results with the prediction of the relativistic
interpretation of brightness asymmetry between the jet and the counterjet.
In this interpretation, the difference in brightness is explained solely by 
the Doppler effect. For a spectrum with a constant spectral index $\alpha$,
one has: $I_j/I_{cj} =
(D_j/D_{cj})^\delta = {\rm D}^\delta$ where $I_j$ ($I_{cj}$) is the jet 
(counterjet) intensity
at a given observational frequency, $D_j$ ($D_{cj}$) is the jet (counterjet)
Doppler factor, and $\delta=3+\alpha$ ($\delta=2+\alpha$ for a filled and 
uniform jet; Blandford and K\"onigl, 1979). 
By assumption in the relativistic interpretation, the physics of the jet and 
the counterjet is the same, so that $\alpha$ is identical for both. 
The intensity ratio between the two jets in the
radio yields ${\rm D}=D_j/D_{cj}= (1+\beta\cos\theta) / (1-\beta\cos\theta)$ 
which is characteristic of the jet and does 
not depend on the frequency of the observations.  
Following Hardcastle et al. (1996), $\alpha=0.5$ between 8 and 15~GHz and  
D is about 2.2 below 7 arcsec from the nucleus and about 1.5 between 7 and 19 
arcsec with $\delta=3+\alpha$ 
(respectively 3.1 and 1.7 with $\delta=2+\alpha$).
The counterjet intensity measured in the I filter then corresponds 
roughly to the intensity of the jet as measured in the U filter (Fig.~5). 
Since the cutoff frequency is in this region of the spectrum, $\alpha$ varies
and the expected intensity ratio between the counterjet and the jet is much 
lower than the same ratio in the radio (Fig.~5). 
We thus have: $I_{cj}(I)\simeq I_j(U)\ (D_j/D_{cj})^{-3}$ or
$\simeq I_j(U)\ (D_j/D_{cj})^{-2}$ depending on the choice for $\delta$. 
Taking the intensity $I_j(U)\simeq 1.1~\mu$Jy of the brightest knot B from 
Fraix-Burnet et al. (1989): $I_{cj}(I)\simeq 0.1~\mu$Jy below
7 arcsec and $\simeq 0.35~\mu$Jy farther away. This is well above the
noise level of the total median image (see Sect. 2). Hence the present data 
should reveal 
the equivalent of knot B for the counterjet, if it is the perfect symmetric of
the jet.

Strictly speaking, there is no symmetric knot to knot B, so that the 
optical emission of the counterjet is fainter than what is predicted by the
relativistic interpretation. But, if the detected knots belong to the 
counterjet, all the `inner' ones
($d<7$~arcsec) are much brighter than the expected $0.1~\mu$Jy. 
The `outer' knots of the jet
(E and F) have U intensities 2 to 3 times lower than knot B, so that the 
expectation in the symmetric region on the counterjet side is rather
$\simeq 0.13$ than $0.35~\mu$Jy as stated above. All corresponding knots are 
significantly brighter than this. As a conclusion, all
identified knots for the counterjet are brighter than the predicted
value from the relativistic interpretation. They even have smaller 
radio-to-optical spectral indices (see Table~1) than the jet, in total
contradiction with the relativistic interpretation (Fig.~5) which predicts
larger spectral indices. 
The values for the spectral index (about 0.5-0.6) are fully consistent with a
synchrotron origin for the optical emission and correspond to values generally
found in jets. The only exception is knot b'
which has a very low spectral index (0.35) so that its link with the radio
jet can be questioned.

The consequence is that the detected knots, if belonging to the counterjet, 
prove
that the radiation properties are intrinsically different in the jet and in 
the counterjet. The counterjet is less bright in the radio implying 
fewer radiating particles as a whole or a lower magnetic field. 
Since the radio-to-optical spectral index is smaller, the cutoff frequency in
the counterjet is higher, which is consistent with a lower magnetic field.

\section{Conclusion}

The search for the first optical extragalactic counterjet has yielded
several components that are coincident with the radio counterjet. 
However, it is 
not possible to determine precisely the origin of the optical emission.
Still deeper images or observations in another filter are needed. From
the radio-to-optical spectral index of these components, it is quite plausible
that they are the optical counterparts of the radio counterjet. If this is the
case, smaller spectral indices are found than for the 
jet, in contradiction with the prediction
by the relativistic beaming interpretation of brightness asymmetry between
the jet and the counterjet. This indeed means that
the physics of the counterjet is intrinsically different from that of the jet:
the magnetic field seems to be weaker in the counterjet, so that the radio
intensity is lower and the cutoff frequency higher. Alternatively, if the
detected knots have no relation with the counterjet, then our data seem to
indicate that its optical emission is fainter than what is predicted by the 
relativistic beaming interpretation, favouring again an intrinsic
brightness asymmetry.

\section*{Acknowledgments}

These observations have been obtained while being visitor astronomer at the 
Canada-France-Hawaii Telescope, operated
by the Conseil National de la Recherche of Canada, the Conseil National de la 
Recherche Scientifique of France and the University of Hawaii. 
I thank M. Cr\'ez\'e, J.-C. Cuillandre, Y. Mellier and A. Robin, for obtaining
some of the data. M.J. Hardcastle very kindly provided 
the electronic radio images.

\bye